\documentclass[12pt]{iopart}
\usepackage{iopams}

\usepackage{amsfonts}
\usepackage{amssymb}
\usepackage{mathrsfs}
\usepackage{graphicx}

\let\vaccent=\v % rename builtin command \v{} to \vaccent{}
\renewcommand{\v}[1]{\ensuremath{\mathbf{#1}}} % for vectors
\newcommand{\gv}[1]{\ensuremath{\mbox{\boldmath$ #1 $}}} % for vectors of Greek letters
\newcommand{\m}[1]{\ensuremath{\mathbf{#1}}}%for matricies
\newcommand{\op}[1]{\ensuremath{\hat{#1}}} % for operators
\newcommand{\abs}[1]{\left| #1 \right|} % for absolute value
\newcommand{\grad}[1]{\gv{\nabla} #1} % for gradient
\newcommand{\vint}[1]{\int\mathrm{d}^3 #1\,}% for volume integral

\def\ii{\imath}
\def\up{\uparrow}
\def\down{\downarrow}

\begin{document}

\title[Time-Reversal-Symmetry-Broken State]{Time-Reversal-Symmetry-Broken State in the BCS Formalism for a Multi-Band Superconductor}

\author{Brendan J. Wilson and Mukunda P. Das}

\address{Department of Theoretical Physics, Research School of Physics and Engineering,
The Australian National University, Canberra ACT 0200, Australia}

\ead{brendan.wilson@anu.edu.au}
\ead{mukunda.das@anu.edu.au}

\begin{abstract}
In three-band BCS superconductors with repulsive interband interactions, frustration between the bands can lead to an inherently complex gap function, arising out of a phase difference between the bands in the range $0$ and $\pi$. Since the complex conjugate of this state is also a solution, the ground state is degenerate, and there appears a time-reversal-symmetry-broken state. In this paper we investigate the existence of this state as a function of interband coupling strength and show how a new phase transition appears between the TRSB and conventional BCS states.
\end{abstract}


\section{Introduction}
Over the past two decades a series of novel superconductors have been discovered bridging the gap between the conventional low $T_c$ superconductors and high $T_c$ cuprates. These materials include Boro-carbides, $\mathrm{MgB}_2$, Pnictides and many others. While the conventional metallic superconductors are adequately explained by the phonon coupled BCS pairing mechanism, now there appears several proposals to address various anomalous features in the novel superconductors. Some are attempted with a modified BCS mechanism (involving both phonon and non-phonon) and others are radically different propositions (Resonance valance bond, marginal Fermi Liquid etc.). The new features of unconventional/novel superconductors involve some of the following: (a) many of them are quasi-2d or 3d materials, (b) superconducting properties are anisotropic, (c) Fermi surfaces are much more complicated involving many bands of electronic structure, (d) a variety of superconducting gap symmetries (singlet or triplet and s-wave, p-wave and d-wave) are confirmed by angle-resolved photo-emission and tunnelling probes and (e) control of superconducting $T_c$ by `doping' into the parent materials.  We avoid these broad range of issues in this paper and focus on one important aspect: how the inter-band coupling in multi-band superconductors can give rise to some novel physics. Before we dwell on the technicality of more exotic states based on the symmetry of the phases, we present a brief appraisal of the multi-band superconductors.

Suhl et al. \cite{Suhl} and Moskalenko \cite{Moskalenko} independently extended the one band BCS theory with overlapping of energy bands on the Fermi surface. The two-band theory was further discussed by Kondo \cite{Kondo} and Tilley \cite{Tilley}, where the solutions are obtained by using Bogoliubov's variational and Gorkov's techniques respectively. Additionally Tilley derived Ginzburg-Landau eqns in the two band case following the method of Gorkov. The two-band model has been explored intensely starting with the work of Geilikman et al. \cite{Geilikman}. Other authors used this model in the context of high $T_c$ oxides \cite{Lee} and MgB$_2$ \cite{Zhitomirsky, Mazin, Kogan,Gurevich}. A comprehensive review of the later topic is presented by Brandt and Das \cite{Brandt}.  

 A new phenomenon was discovered to exist in multi-band superconductors by Leggett \cite{Leggett} (see also \cite{Stanev}), where a collective oscillation was predicted due to small fluctuations in the relative phases of two order parameters. This oscillation appears because the total energy of the system not only is dependent on the relative phases, but also on its canonical conjugate variable, the relative density of the bands.  The inter-band coupling in this case is the internal Josephson coupling usually written as $V_{ij} = g_{ij}  \abs{\Delta_i} \abs{\Delta_j } \cos(\phi_i-\phi_j)$. $ g_{ij}$ is the strength of coupling,  $\abs{\Delta_i}$  and $\phi_i$  are amplitude and phase of the energy gap. Search for Leggett modes have been pursued with no clear success \cite{Ya,Blumberg}.

Conventionally, the gaps $\Delta_i(r,t) =g_{ij}(r,t)\mathscr{F}_j(r,t)$, where $\mathscr{F}_j(r,t)$ are the anomalous Green functions, are taken to be even functions of time.  Their Fourier transforms, therefore, are even functions of frequency, $\omega$. For sufficiently strong exchange interactions Berezinskii \cite{Berezinskii} showed a novel pairing state having triplet spin and even orbital angular momentum, $l =0$. This led to $\mathscr{F}$ and $\Delta$ being odd functions of $t$ and $\omega$, a strangely different property unlike the conventional superconductors \cite{Schrieffer}.  Abrahams et al.\cite{Abrahams} have pointed out that for electron-phonon interaction in metals, the odd frequency state can not exist due to renormalisation of the net interaction, but with anti-ferromagnetic spin fluctuations the odd frequency state seems to be a clear possibility.  Symmetry of the superconducting gap under time reversal for various spin and orbital angular momenta channels give rise to many novel superconducting classes (see Table 1 in \cite{Schrieffer}). Physics of pair tunnelling with different gap parameters in multi-band systems has opened up a new paradigm in the current research.

In three band superconductors, a chiral state appears as a new possibility for a ground state, termed the time-reversal-symmetry-broken (TRSB) state \cite{Agterberg}.  Chiral superconductivity occurs due to the competition of attractive and repulsive inter-band interactions \cite{Tanaka, Hu, Lin,Yanagisawa,Tesanovic}, allowing the inter-band phase differences to take on values other than $0$ and $\pi$. The TRSB state is then doubly degenerate, with the complex conjugate of the ground state having the same energy. In this state, novel arrangements can appear due to variations in this phase difference, including skyrmions\cite{Babaev}, fractional vortices, and phase solitons. 

In this paper we report a modified BCS theory for the multi-band case and consider in particular a 3-band model for pure superconductors.  The superconducting gap amplitudes on the different sheets of the Fermi surface give rise to multi-gap phenomena. Unlike in the two-band case, where the inter-band coupling always enhances the superconducting gaps and the transition temperature, the transition temperature of a three-band superconductor can be enhanced or reduced,  depending on the sign and magnitude of inter-band couplings. Because of multi-band gaps (amplitudes and phases) being different, a variety of novel phenomena are expected. Some of these details constitute the core of this paper.  In Sec 2 we outline the multi-band BCS theory and show numerical solution of the gaps as a function of $T/T_c$. In Sec 3 we discuss time reversal broken symmetry state and the phase diagram of three band superconductors. In Sec 4. our results and a summary is presented.

\section{Multi-Band BCS Theory}

To generalise the BCS theory to multi-band superconductors, a band index $i$  is added to the electron operators and the bands are coupled by an interband Josephson coupling term. The Hamiltonian is written in momentum space as
\begin{eqnarray}
\op{H}=&\sum_{\v{k}i\sigma}(\epsilon_{i\sigma}(\v{k})-\mu)\op{c}_{i\sigma}^\dag(\v{k})\op{c}_{i\sigma}(\v{k})\nonumber\\
&+\sum_{\v{kk'}ij\sigma\sigma'}V_{ij}(\v{k},\v{k'})\op{c}_{j\sigma}^\dag(\v{k})\op{c}_{j\sigma'}^\dag(\v{k'})\op{c}_{i\sigma'}(\v{k'})\op{c}_{i\sigma}(\v{k}),
\end{eqnarray}
where $\epsilon_{i\sigma}(\v{k})$ is the energy of the non interacting state, $\sigma=\up,\down$ indicates the spin state, $i,j=1,2$ is the band index, $\op{c}_{i\sigma}(\v{k})$ ($\op{c}\dag_{i\sigma}(\v{k})$) are the electron annihilation (creation) operators in momentum space, $\mu$ is the chemical potential, $V_{ij}(\v{k},\v{k'})$ is the general momentum dependent intraband ($i=j$) and interband ($i\neq j$) coupling strengths. The Hamiltonian is also written in real space as
\begin{eqnarray}
\op{H}=&\sum_{i,j}g_{ij}\vint{x} \op{\psi}^\dag_{j\up}(\v{x})\op{\psi}^\dag_{j\down}(\v{x})\op{\psi}_{i\down}(\v{x})\op{\psi}_{i\up}(\v{x})+\nonumber\\
&\sum_{i\sigma}\vint{x} \op{\psi}^\dag_{i\sigma}(\v{x})\left(\frac{1}{2m_i}\left(-\ii\hbar\grad-\frac{e\v{A}(\v{x})}{c}\right)^2-\mu\right)\op{\psi}_{i\sigma}(\v{x}),
\end{eqnarray}
where $\op{\psi}_{j\up}(\v{x})$ ($\op{\psi}\dag_{j\up}(\v{x})$) are the electron annihilation (creation) operators in real space, $\v{A}(\v{x})$ is the vector potential, $m_i$ is the effective mass of the electron in the band $i$, $g_{i j}$ is the effective interband coupling constant, and it is implicitly assumed that all real space integrals must be cut off in momentum space at the Debye frequency, $\omega_D$.

In the absence of an external magnetic field, the superconducting gap satisfies the self-consistent gap equations, which are given by
\begin{eqnarray}
\m{g}^{-1}\gv{\Delta}&=\m{L}(\v{\Delta},T)\v{\Delta},
\label{eq:MultibandBCSVectorEqn}
\end{eqnarray}
where $\gv{\Delta}=\{\Delta_1,...,\Delta_N\}$ is a vector containing the gap in each band, $\m{g}$ is the effective interband coupling matrix, and $\m{L}(\v{\Delta},T)$ is a diagonal matrix with elements
\begin{eqnarray}
\m{L}(\v{\Delta},T)_{jj}&=L(\Delta_j,T)\nonumber\\
&=N_{j}(0)\int_0^{\hbar\omega_D}\frac{\mbox{ d}\xi}{\sqrt{\xi^2+\abs{\Delta_j}^2}}\tanh\left(\frac{\sqrt{\xi^2+\abs{\Delta_j}^2}}{2k_BT}\right).
\end{eqnarray}
Here $N_j(0)$ is the density of states of each band at the Fermi surface.
The critical temperature, $T_c$, occurs when there is a non trivial solution to these equations. This temperature is found as the largest temperature when the following determinant is zero
\begin{equation}
\det\left[\m{g}^{-1}-\m{L}(\v{0},T_c)\right]=0,
\end{equation}
where now the integral is the same for all bands with the approximate solution
\begin{eqnarray}
\eta&=\int_0^{\hbar\omega_D}\frac{\mbox{ d}\xi}{\xi}\tanh\left(\frac{\xi}{2k_BT_c}\right)\nonumber\\
&=\ln\left(\frac{2\hbar\omega_D e^\gamma}{\pi k_BT_c}\right).
\end{eqnarray}
Here $\gamma\approx 0.577216$ is the Euler-Mascheroni constant and $\omega_D$ is the Debye frequency.

For the three band BCS, $T_c$ is the largest solution to the equation
\begin{equation}
0=\left|
\begin{array}{ccc}
g_{11}^{-1}-\eta N_{1}(0) 	&  g_{12}^{-1} 			&  g_{13}^{-1} \\
g_{21}^{-1}				&  g_{22}^{-1}-\eta N_{2}(0)	&  g_{23}^{-1} \\		
g_{31}^{-1}		 		&  g_{32}^{-1}			&  g_{33}^{-1}-\eta N_{3}(0)
\end{array}
\right|.
\end{equation}

We can also calculate $\Delta_0$, the gap at zero temperature. In this limit the integral can be performed exactly with the result
\begin{eqnarray}
\int_0^{\hbar\omega_D}\frac{\mbox{ d}\xi}{\sqrt{\xi^2+\Delta_0^2}} &=\log\left(\frac{\hbar\omega_D+\sqrt{\hbar\omega_D^2+\abs{\Delta_0}^2}}{\abs{\Delta_0}}\right)\\
&\approx \log\left(\frac{2\hbar\omega_D}{\abs{\Delta_0}}\right).
\end{eqnarray}

Then the gap at $T=0$ is found by solving the set of equations
\begin{eqnarray}
\sum_i g_{ij}^{-1}\Delta_{0i}&=N_j(0)\Delta_{0j}\log\left(\frac{\hbar\omega_D+\sqrt{\hbar\omega_D^2+\abs{\Delta_{0j}}^2}}{\abs{\Delta_{0j}}}\right).
\end{eqnarray}

\begin{figure}
\centering
\includegraphics[width=0.5\textwidth]{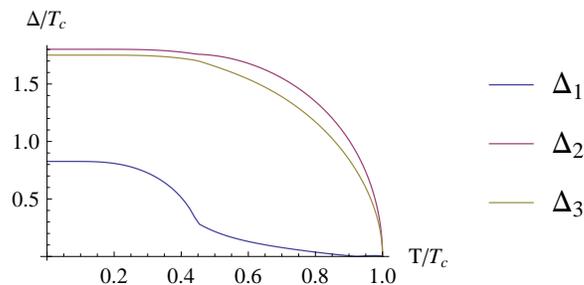}
\caption{(Colour online) This shows a numerical solution to the three-band BCS equations. Notice the kink at $T/T_c\approx0.45$.  Below this point the superconductor is in the time-reversal-symmetry-broken state.}
\label{fig:ThreeBandWithTStar3Gap}
\end{figure}

Numerical solutions for a three band case over all $T$ are shown in Fig ~(\ref{fig:ThreeBandWithTStar3Gap}). For this plot we use the parameters: $\hbar\omega_D=0.9$, $g_{11}=0.7$, $g_{22}=0.75$, $g_{33}=0.6$, $g_{12}=g_{13}=g_{23}=-0.1$, $N_1(0)=0.4$, $N_2(0)=0.5$, $N_3(0)=0.6$. With these parameters, a kink appears in the gap at about $T/T_c=0.45$. This kink is more pronounced in the smallest gap, but is present in all three bands. Above this temperature, the phase difference between the gaps must either be $0$ or $\pi$. In this region the superconductor is in a conventional BCS state. Below this point $(T/T_c<0.45)$ the phase difference will be between $0$ and $\pi$, and the superconductor is in the TRSB state (Also see \cite{Dias,Tanaka2}).

\section{Calculation of the TRSB State}

\begin{figure}[t!]
\centering
\includegraphics[width=0.5\textwidth]{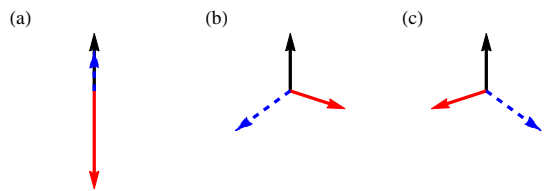}
\caption{(Colour online) (a) For most ground state solutions, all bands are in phase or out of phase with each other. After removing an arbitrary phase, the gap at a point can be described by a real number. (b)-(c) In the time-reversal-symmetry-broken state, the states can have any phase between them. In this case, after removal of the an arbitrary phase, the gaps are still described by a complex number. The ground state is thus doubly degenerate, with the ground states being complex conjugates of each other.}
\label{fig:arrow}
\end{figure}

For a three-band superconductor, most solutions of the self-consistent multi-band BCS equations have the three gaps either in phase or out of phase with each other. However, for some input parameters and below a critical temperature $T^*$, there exists a solution to the BCS  self-consistent equations (Eq \ref{eq:MultibandBCSVectorEqn}) where the three bands have relative phases other than $0$ or $\pi$ due to frustration of the interaction between the three gap functions. These gap functions are therefore described by a complex amplitude even in the uniform case. When this happens the ground state is doubly degenerate because the complex conjugate of the gap functions is also a solution to Eq (\ref{eq:MultibandBCSVectorEqn}). The gap functions are therefore in a chiral ground state, and the time-reversal symmetry is spontaneously broken.

Considering the gap functions as vectors in the complex plane (See Fig ~(\ref{fig:arrow})), we see that the two possible ground states have different chirality. This induces a breaking of the mirror symmetry in the phase plane, which also induces the breaking of the time-reversal-symmetry.

Here we show how to calculate the magnitude of the gap in the TRSB state in terms of a set of single band BCS gap equations, and find the condition for the appearance of the TRSB state.

For the three-band case, the self-consistent BCS equations (Eq (\ref{eq:MultibandBCSVectorEqn})) can be written in matrix form as
%\begin{widetext}
\begin{eqnarray}
\label{eq:ThreeBandSelfConsistantMatrixForm}
\left[
\begin{array}{ccc}
L(\Delta_1,T)-g^{-1}_{11} 	& -g^{-1}_{21}				& -g^{-1}_{31} \\
-g^{-1}_{12} 				& L(\Delta_2,T)-g^{-1}_{22} 	& -g^{-1}_{32} \\
-g^{-1}_{13} 				& -g^{-1}_{23}				& L(\Delta_3,T)-g^{-1}_{33} \end{array}
\right]
\left[
\begin{array}{c}
\Delta_1\\
\Delta_2\\
\Delta_3 \end{array}
\right]=\v{0},
\end{eqnarray}
%\end{widetext}
where there is an arbitrary global phase. We can choose the global phase such that $\Delta_1=\abs{\Delta_1}$.
Then, taking the imaginary part, we find
\begin{eqnarray}
\left[
\begin{array}{cc}
L(\Delta_2,T)-g^{-1}_{22} 	& -g^{-1}_{32} \\
-g^{-1}_{23}				& L(\Delta_3,T)-g^{-1}_{33} \end{array}
\right]
\left[
\begin{array}{c}
\mathrm{Im}(\Delta_2)\\
\mathrm{Im}(\Delta_3) \end{array}
\right]=\v{0}.
\end{eqnarray}

\begin{figure}[t!]
\centering
\includegraphics[width=0.5\textwidth]{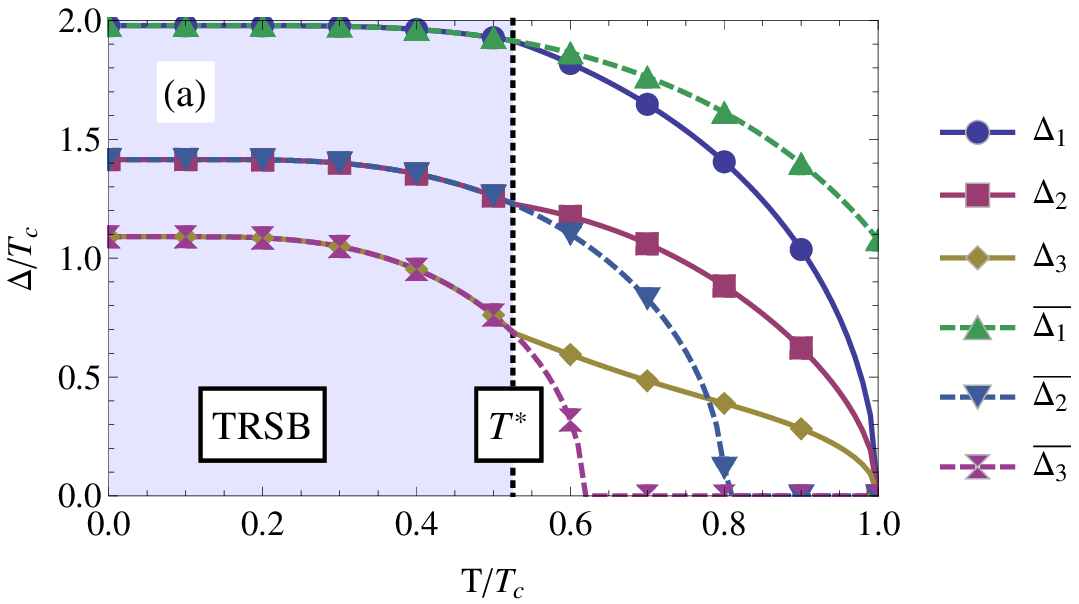}\\
\includegraphics[width=0.5\textwidth]{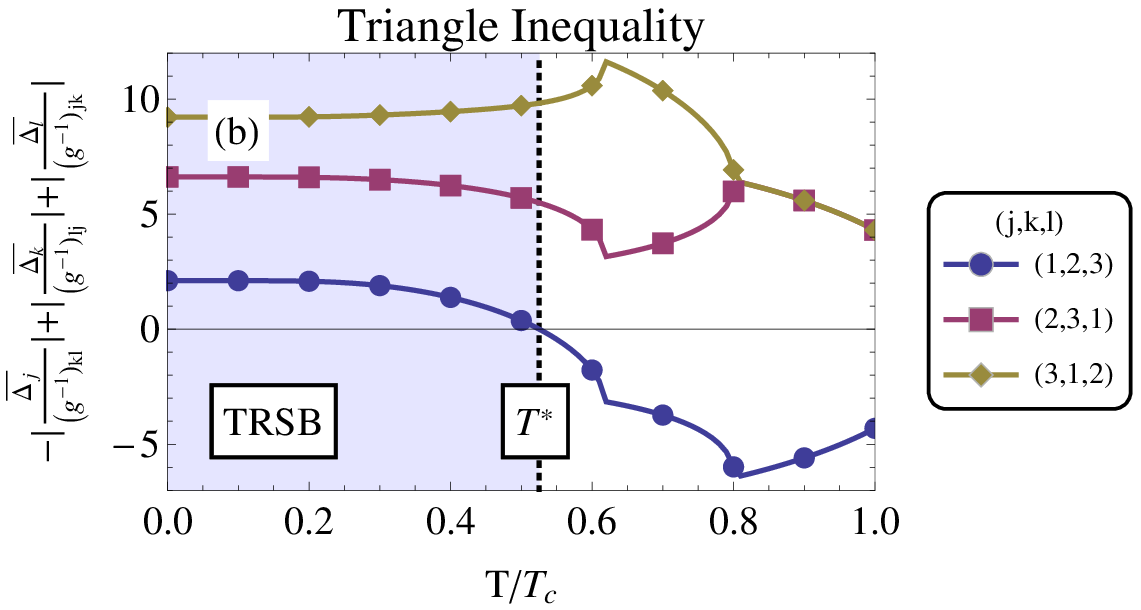}
\caption{(Colour online) a) This shows a direct solution to the three-band BCS equations as the three solid lines, as well as the one band TRSB states shown with dashed lines. In the TRSB state the multi-band solution agrees with a reduced one-band solution. b) The triangle inequality for the scaled reduced one-band gaps is plotted as a function of temperature. The point where the triangle equality fails is the same location as where the one-band solution departs from the multi-band solution in the previous plot, $T^*$.}
\label{fig:TStarConditions}
\end{figure}

If we are interested in the time-reversal symmetry broken states, then the imaginary components of these gaps must be non-zero. Hence the determinant of the matrix must equal zero. We will denote a magnitude of the gap that satisfies this requirement $\overline{\Delta_i}$. This leads to the constraint equation
\begin{eqnarray}
(L(\overline{\Delta_2},T)-g^{-1}_{22})(L(\overline{\Delta_3},T)-g^{-1}_{33})-(g^{-1}_{23})^2=0.
\end{eqnarray}
We can make similar demands for the other gaps, resulting in two other similar conditions. These have the solution
\begin{eqnarray}
\label{eq:onebandBCScoupling}
L(\overline{\Delta_j},T)&=g^{-1}_{jj}\pm\frac{g^{-1}_{jk}g^{-1}_{jl}}{g^{-1}_{kl}},
\end{eqnarray}
for all permutations $j,k,l=1,2,3$, which are now decoupled single-band BCS self-consistent equations for the superconducting gaps with modified interaction strengths.

\begin{figure}[t!]
\centering
\includegraphics[width=0.5\textwidth]{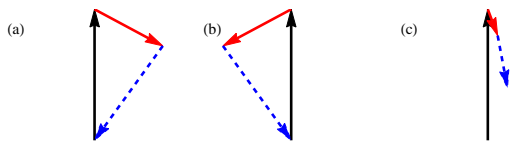}
\caption{(Colour online) The triangle inequality can be viewed as a restriction on being able to form a triangle with three vectors of a given length. (a) \&(b). If the triangle inequality is satisfied, then two triangles can be formed which are mirror images of each other. (c) If the triangle inequality is not satisfied, then it is not possible to form a triangle with the vectors of that length because one of the vectors is too long.}
\label{fig:TriangleInequalityArrow}
\end{figure}

Since the TRSB state is only stable when one or all three coupling parameters are negative, $g_{12}g_{13}g_{23}<0$, resulting in frustrated repulsive interactions between the bands, we only need to keep the negative solution. We can now substitute Eq.~(\ref{eq:onebandBCScoupling}) into Eq.~(\ref{eq:ThreeBandSelfConsistantMatrixForm}) to find
\begin{eqnarray}
\label{eq:ThreeBandSelfConsistantMatrixForm2}
\left[
\begin{array}{ccc}
-\frac{g^{-1}_{12}g^{-1}_{13}}{g^{-1}_{23}} 	& -g^{-1}_{21}						& -g^{-1}_{31} \\
-g^{-1}_{12} 						&-\frac{g^{-1}_{12}g^{-1}_{23}}{g^{-1}_{13}}  	& -g^{-1}_{32} \\
-g^{-1}_{13} 						& -g^{-1}_{23}						& -\frac{g^{-1}_{13}g^{-1}_{23}}{g^{-1}_{12}} 
 \end{array}
\right]
\left[
\begin{array}{c}
\overline{\Delta_1}\\
\overline{\Delta_2}\\
\overline{\Delta_3} \end{array}
\right]=0,
\end{eqnarray}
which results in a single condition which must be met for a stable solution. This can be rewritten in the form
\begin{eqnarray}
\frac{\overline{\Delta_1}}{g^{-1}_{23}}+\frac{\overline{\Delta_2}}{g^{-1}_{13}}+\frac{\overline{\Delta_3}}{g^{-1}_{12}}=0,
\end{eqnarray}
where the $\overline{\Delta_i}$ are complex with unknown phase differences. The existence of a solution to this constraint equation is only possible if the triangle inequality
\begin{eqnarray}
\label{eq:TriangleInequality}
\abs{\frac{\overline{\Delta_j}}{g^{-1}_{kl}}}\leq\abs{\frac{\overline{\Delta_k}}{g^{-1}_{lj}}}+\abs{\frac{\overline{\Delta_l}}{g^{-1}_{jk}}},
\end{eqnarray}
holds for all permutations of $\{j,k,l\}$. In this way we can find the existence and position of the time-reversal-symmetry breaking by finding the solution to  Eq.~(\ref{eq:onebandBCScoupling}), then finding the regions over which Eq. (\ref{eq:TriangleInequality}) holds. 

In Fig.~(\ref{fig:TStarConditions}) we perform this procedure explicitly with the parameters $\hbar\omega_D=0.09$, $g_{11}=g_{22}=g_{33}=0.7$, $g_{12}=g_{13}=g_{23}=-0.1$, $N_1(0)=0.35$, $N_2(0)=0.32$, $N_3(0)=0.3$. This set of parameters are arbitrary, yet reasonable for a three band case. Similar results are found with other parameters. Figure (a) shows a direct multi-band solution to the BCS equations, $\Delta$, plotted with the one-band reduced BCS solutions, $\overline{\Delta}$. In figure (b) the triangle inequality for the three permutations are plotted as a function of temperature. These plots show that in the region where the inequalities hold, the multi-band solution is precisely the solution $\overline{\Delta}$. At the point where the inequality is no longer satisfied there is a kink in the multi-band solution as it ceases to be in the TRSB state. The point where the equality holds is the critical transition point $T^*$, and is the point that separates the TRSB state from the conventional BCS state.

At the point $T^*$, the full multi-gap solution separates from the reduced one band solution with a discontinuity in the derivative of the gaps. This discontinuity would lead to a jump in the heat capacity of the superconductor and other experimentally measurable quantities.

Looking at the TRSB triangle inequality (Eq (\ref{eq:TriangleInequality})), we see that there can be points above $T^*$ where the permutation below zero changes sign, and another permutation drops below zero. At the point where they cross, the triangle inequality holds as an equality for the two permutations, and therefore the solution is again the reduced one-band BCS solution. As one of the one-band solutions is zero at this point, this allows is the phase of this gap to change relative to the other two. We denote this point $T^0$, and note that this point can be seen most readily as a phase discontinuity.

\section{Results and Summary}

\begin{figure*}[ht!]
\centering
\includegraphics[width=0.7\textwidth]{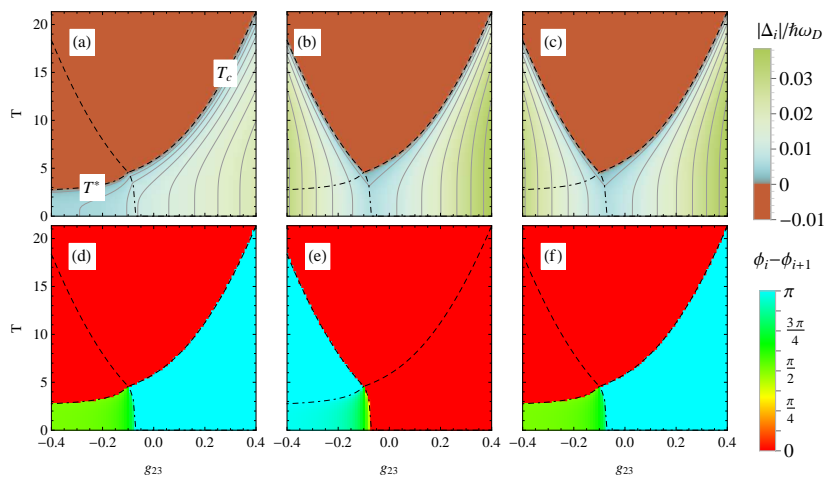}
\caption{(Colour online) Phase Diagram with equal bands. $\hbar\omega_D=0.09$, $N_i(0)=0.2$, $g_{ii}=0.8$, $g_{12}=g_{13}=-0.1$. (a)-(c) The gaps in bands 1-3, $\abs{\Delta_i}$, respectively. (d)-(f) The phase difference between the gaps. (d) $\phi_1-\phi_2$.  (e) $\phi_2-\phi_3$.  (f) $\phi_3-\phi_1$.}
\label{fig:EqualBandPhaseDiagram}
\end{figure*}

\begin{figure*}[ht!]
\centering
\includegraphics[width=0.7\textwidth]{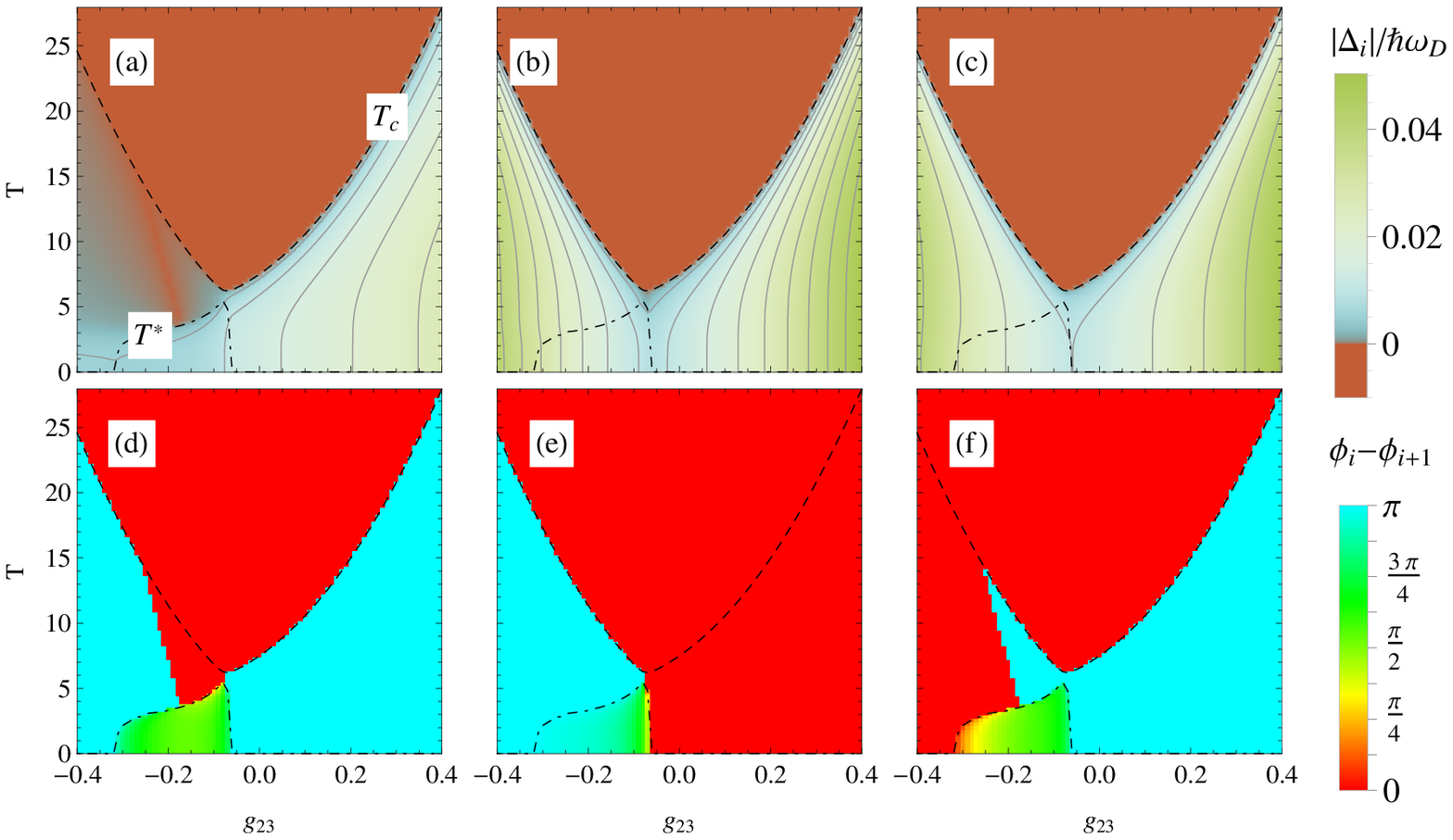}
\caption{(Colour online) Phase Diagram with nearly equal bands showing two phase discontinuities. $\hbar\omega_D=0.09$, $N_1(0)=0.2$, $N_2(0)=0.21$, $N_3(0)=0.22$, $g_{ii}=0.8$, $g_{12}=-0.11$, $g_{13}=-0.1$. (a)-(c) The gaps in bands 1-3, $\abs{\Delta_i}$, respectively. (d)-(f) The phase difference between the gaps. (d) $\phi_1-\phi_2$.  (e) $\phi_2-\phi_3$.  (f) $\phi_3-\phi_1$.}
\label{fig:NearlyEqualBandPhaseDiagram}
\end{figure*}

\begin{figure*}[ht!]
\centering
\includegraphics[width=0.7\textwidth]{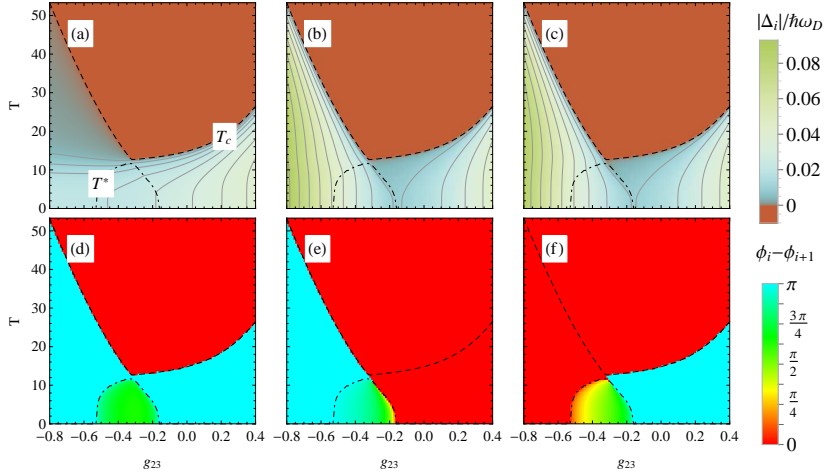}
\caption{(Colour online) Phase Diagram with nearly equal bands with one phase discontinuity. $\hbar\omega_D=0.09$, $N_1(0)=0.3$, $N_2(0)=0.21$, $N_3(0)=0.22$, $g_{ii}=0.7$, $g_{12}=-0.11$, $g_{13}=-0.1$. (a)-(c) The gaps in bands 1-3, $\abs{\Delta_i}$, respectively. (d)-(f) The phase difference between the gaps. (d) $\phi_1-\phi_2$.  (e) $\phi_2-\phi_3$.  (f) $\phi_3-\phi_1$.}
\label{fig:OneT0LinePhaseDiagram}
\end{figure*}

With the choice of a base set of parameters, we can vary a single parameter and observe the effect on the magnitude and relative phase of the gaps, as well as on the critical temperature $T_c$, the presence of the TRSB state, and the existence of a $T^0$ point.

We first choose the case where the parameters of all the bands are equal other than the parameter $g_{23}$ (See Fig. (\ref{fig:EqualBandPhaseDiagram})). We choose the parameters $\hbar\omega_D=0.09$, $g_{11}=g_{22}=g_{33}=0.8$, $g_{12}=g_{13}=-0.1$, $N_1(0)=N_2(0)=N_3(0)=0.2$. In this case the parameter space divides into four regimes: at the top is the normal state, on the right is the simple three band superconductivity, bottom left is the TRSB state, and at the top left the first band is identically zero, so that two band superconductivity remains here. These four regions are separated by the phase transition lines $T_c$ and $T^*$, which touch at the point $g_{23}=g_{12}=g_{13}$, which coincides with a minimum in $T_c$.

Due to bands 2 and 3 being identical, we see that the TRSB state continues to exist to $g_{23}=-\infty$, since any non-zero value for $\abs{\Delta_1}$ will satisfy the triangle inequality. Moreover, when $\abs{\Delta_1}=0$ the triangle inequality is still satisfied as an equality, and thus the condition for $T^0$ is satisfied in the entire region above $T^*$ left of the point $T^*=T_c$. Therefore the gap in the first band $\abs{\Delta_1}$ equals zero in this region, and the phase difference for bands $2$ and $3$ is flipped to account for the repulsive interaction between these bands. 

Next we look at the case where the parameters in all three bands are nearly equal (See Fig. (\ref{fig:NearlyEqualBandPhaseDiagram})). Here we set $\hbar\omega_D=0.09$, $g_{11}=g_{22}=g_{33}=0.8$, $g_{12}=-0.11$, $g_{13}=-0.1$, $N_1(0)=0.2$, $N_2(0)=0.21$, $N_3(0)=0.22$. With these parameters the two-band superconductivity regime disappears, the TRSB regime is confined to a finite region, and a gap opens between the $T^*$ line and the $T_c$ line. 

The two-band superconductivity regime disappears due to the finite difference between the bands $\Delta_2$ and $\Delta_3$. Therefore, as the temperature is increased, the $\overline{\Delta_1}$ term eventually becomes too small to satisfy the triangle inequality, and is therefore no longer forced to zero for the majority of this region. However, it remains heavily suppressed.

The finite region containing the TRSB state is also a consequence of the finite difference between the bands $\Delta_2$ and $\Delta_3$. As the coupling $g_{23}$ is moved to $-\infty$, it enhances $\overline{\Delta_2}$ and $\overline{\Delta_3}$ equally, and hence the difference between them, while $\overline{\Delta_1}$ approaches a constant. Therefore, at some critical value of $g_{23}$, the magnitude of $\overline{\Delta_1}$ is too small to satisfy the triangle inequality, even at $T=0$, and so the TRSB state disappears.

The gap opens between the $T^*$ line and the $T_c$ line because for these two points to touch there must be a location where all the terms in the triangle inequality are identical for all temperatures such that the TRSB state can extend all the way to $T_c$. However, the $T_c$ and $T^*$ lines are connected by two $T^0$ lines, one where $\Delta_1=0$, and one where $\Delta_2=0$. These lines are most easily seen in the phase plots as a sudden jump from a phase difference of $0$ to $\pi$ or vice versa, but the line for $\Delta_1$ is also visible in the magnitude plot. We also note that the TRSB state is mainly found near the minimum in $T_c$. 

Finally we look for a set of parameters that results in only a single $T^0$ line. We find that the parameters $\hbar\omega_D=0.09$, $g_{11}=g_{22}=g_{33}=0.7$, $g_{12}=-0.11$, $g_{13}=-0.1$, $N_1(0)=0.3$, $N_2(0)=0.21$, $N_3(0)=0.22$ create this behaviour (See Fig. ~(\ref{fig:OneT0LinePhaseDiagram})). The $T^0$ line is again most easily seen in the phase plots, which show that in this case it is $\Delta_3$ which goes to zero in this case.

The first gap is again suppressed in the upper left region, as was the case in the previous examples. The TRSB state again appears only near the minimum in $T_c$.

In summary, we have looked at the novel TRSB state in three-band superconductors. This state is a result of frustrated repulsive interactions among the three bands which gives rise to a relative phase appearing between the bands. We have shown the behaviour of the gaps as a function of temperature and showed that there is a kink in the band gaps at the critical temperature $T^*$. Below $T^*$, the gaps are described by a set of single-band BCS equations, but above this point the full multi-band BCS equations must be solved directly. 

We have also looked at the behaviour of three-band superconductors as the inter-band coupling is varied. This showed that the TRSB state only appears in a very small region of parameter space, where the interactions between the bands are of similar order. This allows the frustrated repulsive interaction to form the TRSB state. The frustrated repulsion  also results in a lowering of the superconducting critical temperature $T_c$, and thus we expect to find these interesting properties near the minimum of $T_c$ in parameter space.

Finally, we have shown that there is a possible phase transition between the TRSB and conventional BCS states at a finite temperature which can be experimentally probed.

\section*{References}{}

%%%%%%%%%%%%%%% BIBLIOGRAPHY%%%%%%%%%%%%%%%%%%


\begin{thebibliography}{99}


\bibitem{Suhl}  H. Suhl, B. T. Matthias and L. R. Walker, Phys. Rev. Lett. \textbf{3}, 552 (1959).
\bibitem{Moskalenko}  V. A. Moskalenko, Fiz. Metal. Metalloved. \textbf{8}, 503 (1959)
\bibitem{Kondo} J. Kondo, Prog. Theo. Phys. \textbf{29}, 1 (1963)
\bibitem{Tilley} D. R. Tilley, Proc. Phys. Soc. \textbf{84}, 573 (1964)
\bibitem{Geilikman} B. T. Geilikman, R. O. Zaitsev and V. Z. Kresin, Sov. Phys.: Solid State \textbf{9}, 642 (1967)
\bibitem{Lee} D. H. Lee and J. Ihm, Solid State Commun. \textbf{62}, 811 (1987).
\bibitem{Zhitomirsky} M. E. Zhitomirsky and V.-H.Dao, Phys. Rev. B \textbf{69}, 054508 (2004)
\bibitem{Mazin} I. I. Mazin and J. Schmalian, Phys. C. Supercond. \textbf{469}, 614 (1995)
\bibitem{Kogan} V. G. Kogan, C. Martin and R. Prozorov, Phys. Rev. B \textbf{80}, 014507 (2009)
\bibitem{Gurevich} A. Gurevich, Phys. Rev. B \textbf{67}, 184515 (2003)
\bibitem{Brandt} E. H. Brandt and M. P. Das, J Supercond. Nov. Magn. \textbf{24}, 57 (2011)
\bibitem{Leggett} A. J. Leggett,  Prog. Theo. Phys. \textbf{36}, 901 (1966)
\bibitem{Stanev} V. Stanev, Phys. Rev. B \textbf{85} 174520 (2012)
\bibitem{Ya}  Ya. G. Ponomarev, S. A. Kuzmichev, M. G. Mikheev, M. V. Sudakova, S. N. Tchesnokov, N. Z. Timergaleev, A. V. Yarigin, E. G. Maksimov, S. I. Krasnosvobodtsev, A. V. Varlashkin, M. A. Hein, G. M\"uller, H. Piel,
L. G. Sevastyanova, O. V. Kravchenko, K. P. Burdina and B. M. Bulychev, Solid State Commun. \textbf{129}, 85 (2004)%arxiv.org/pdf/cond-mat/0303640
\bibitem{Blumberg} G. Blumberg, A. Mialitsin, B. S. Dennis, M.V. Klein, N.D. Zhigadlo and J.Karpinski,  Phys. Rev. Lett. \textbf{99}, 227002 (2007).
\bibitem{Berezinskii} V. L. Berezinskii, JETP Lett. \textbf{20}, 287 (1974)
\bibitem{Schrieffer}  J. R. Schrieffer, A. V. Balatsky, E. Abrahams and D. J. Scalapino, J. Supercond. \textbf{7}, 501 (1994)
\bibitem{Abrahams}  E. Abrahams, A. V. Balatsky, J. R. Schrieffer and P. B. Allen, Phys. Rev. B \textbf{47}, 513 (1993)
\bibitem{Agterberg} D. F. Agterberg, V. Barzykin and L. P. Gor'kov, Phys. Rev. B \textbf{60}, 20 (1999). These authors consider degenerate bands to obtain frustrated superconductivity while our method achieves frustrated superconductivity as a consequence of repulsive interactions.
\bibitem{Tanaka}  Y. Tanaka and T. Yanagisawa, Solid State Commun. \textbf{150}, 1980 (2010)
\bibitem{Hu} X. Hu and Z. Wang, Phys. Rev. B \textbf{85}, 064516, (2012)
\bibitem{Lin} S. Lin and X.Hu, Phys. Rev. Lett. \textbf{108}, 177005 (2012)
\bibitem{Yanagisawa} T. Yanagisawa, Y. Tanaka, I. Hase and K. Yamaji, J. Phys. Soc. Jap. \textbf{81}, 024712 (2012)
\bibitem{Tesanovic} V. Stanev and Z. Te\vaccent{s}anovic, Phys. Rev. B \textbf{81} 134522 (2010)
\bibitem{Babaev}J. Garaud, J. Carlstr\"om, E. Babaev and M. Speight, Phys. Rev. B, \textbf{87}, 014507, (2013)
\bibitem{Dias} R. G. Dias and A. M. Marques, Supercond. Sci. Technol. \textbf{24} 085009 (2011)
\bibitem{Tanaka2} Y. Tanaka J. Phys. Soc. Jpn. \textbf{70}, 10 (2001)

\end{thebibliography}
\end{document}